# Our climate fix must be big and quick.

William H. Calvin*

**A very large carbon sink could be created by sinking sewage in offshore dead zones. Half of all people live close enough to a coast to personally contribute, making the sink size close to half the annual food production.**

Large amounts of carbon have now been removed from long-term storage and added to the air as $CO_2$. While fossil fuel emissions are what has gotten us into climate trouble, it does not follow that fixing them will get us out.

Reduction in the emissions per year only slows the rate at which $CO_2$ concentration rises and further heats the earth. Emissions reduction does nothing to reverse either ocean acidification or our present climate problems (heat waves, drought, and the ten-fold increase since 1950 in the number of major wildfires and floods[1]). It is not until we begin reducing $CO_2$ concentration per se that we are likely to see any improvement. Until then, climate problems will only get worse.

To reduce atmospheric $CO_2$ requires sinking excess carbon. Framing our climate response in terms of gradual emissions reduction has a tendency to make us think that new carbon sinks are a mere contingency plan should emissions reduction fail ("Techniques for extracting atmospheric $CO_2$ …might eventually prove necessary[2]") rather than a parallel approach.

We need enough new carbon sinks to 1) cancel out any continuing use of fossil fuels, 2) overcome the delayed effect of earlier excesses[3], and then 3) lower atmospheric $CO_2$ concentrations to the old maximum value of 280 ppm. We need to sink about 25 billion tonnes of carbon each year, half to counter current emissions and commitments, and half for drawing down the $CO_2$ concentrations within a few decades.

A frame of gradual climate change and gradual response also leaves us open to being blindsided by abrupt events and tipping points[4]. Some abrupt climate change has already occurred. The global land acreage in the two most extreme stages of drought[5] oscillated around 14 percent from 1950 to 1982. With the 1982 mega El Niño, the baseline popped up to 24 percent and stayed there[6] until the 1997 mega El Niño when another baseline shift took drought up to 34 percent of global land surface. Only in 2005 did global drought step down into the high 20s.

Furthermore, we almost had a $CO_2$ jump in 1999. Both the Amazon and Southeast Asian rain forests dried enough in 1997-1998 so that large areas burned[7]. Had this mega El Niño lasted twice as long and caused most of the trees to decompose or burn, there could have been a mass extinction of species and a 40 ppm increase in worldwide $CO_2$. The annual $CO_2$ increment would rise 50 percent—even with constant emissions—because of the loss of leaves.

With a Big Burn causing a 50 percent jump in the excess $CO_2$ (that beyond 280 ppm) in as little as two years, the climate impacts would likely be even greater. Because some climate mechanisms are likely sensitive to how fast temperature rises, not just the amount of overheating, the chance of a climate surprise would increase.

The next mega El Niño[8] (recent ones have been spaced at 10 and 15 year intervals) will be a suspenseful time. The "burn locally, crash globally" scenario illustrates a present danger, not merely a midcentury one[9].

Not only must our carbon sink be big enough but it and emissions reduction must together be fast enough to remove us from the danger zone for abrupt climate change. Risk will depend on how many years we are forced to dwell in such danger zones before backing out of them. A serious jolt would create mass migrations across borders with the attendant famine, pestilence, war, and genocide.

Our ability to avoid a human population crash would be severely compromised by disorganized economies and ineffective international cooperation. To achieve a safety factor, we must front-load our climate response (much as a course of antibiotics may include a double dose the first day).

What to do? For our climate crisis, anthropogenic carbon sinks must be big, quick, and secure. Since the climate forecast is for even more drought and for higher winds to spread fires, planting more trees will not be sufficiently secure. But there is at least one way to meet all three requirements.

First note that it is not necessary to remove $CO_2$ directly from the atmosphere in order to reduce its concentration. Photosynthesis already removes large amounts of $CO_2$ and, at the other end of the carbon cycle, respiration and cell decomposition (combustion, rotting) release $CO_2$. It would suffice to keep atmosphere-bound carbon from reaching its destination. Except to collectively hold our breath, we cannot do anything about respiration's $CO_2$ contribution. Fortunately, anoxic burial of biomass can take carbon out of circulation and lower the atmospheric $CO_2$ concentration.

In some places, the seafloor is an anoxic dead zone because the rain of decomposing plankton from the surface has already used up the oxygen, it is too dark for photosynthesis to produce more, and circulation does not deliver enough from elsewhere. The entire floor of the Black Sea is a natural dead zone, as are many sea floor basins and some areas of the continental shelf[10]. About 80 percent of the world's oil originates from the carbon-rich sediments formed at such anoxic bottoms.

Though anthropogenic dead zones have been appearing offshore of many industrialized countries, wave action will often oxygenate near-shore waters where sewage outfalls are located. Most coastal cities[11] pipe raw or lightly treated sewage just far enough away to minimize odors and fecal contamination of nearby beaches, with outfalls less than 5 km offshore in depths less than 50 m. Thus, ocean sewage sediments often form where there is enough oxygen for making $CO_2$, allowing worms to stir the sediments and fish to spread pathogens.

One solution is to extend existing outfalls into dead zones farther offshore. Alternatively, in cases where circulation is slow, a hypoxic area could be made into a year-round dead zone by locally augmenting surface productivity (adding iron is the familiar example[12]). Either method would result in sewage sediments forming without releasing $CO_2$, thus taking carbon out of the carbon cycle.

Would this annually sequester enough carbon to actually reduce $CO_2$ concentration, so that we can begin backing out of the danger zone for abrupt climate change? This is not the place for that calculation but the amount of sinkable biomass is surely very large, as conservation of mass requires that global production of human excrement must be almost as large as our intake, global food production[13].

About half of the world's population resides close enough to an ocean to contribute to an anthropogenic carbon sink.[14] Inland waste streams such as sewage, feedlot waste, and sawdust could be transported and added to coastal sewage streams; if minced, garbage and crop residue might also be added. Environmental concerns such as $CH_4$ or $H_2S$ release will require evaluation; note that some industrial wastes could be buried by the sewage if the waste streams were combined, allowing troublesome outfalls to be shut down.

For addressing our climate crisis, nothing currently on the table comes close to the usefulness of sequestering sewage on an anoxic seafloor. It is very big, relatively secure, and quickly done.

* University of Washington, Program on Climate Change and School of Medicine, Box 351800, Seattle WA 98195-1800 USA. E-mail WCalvin@U.Washington.edu.

[Editors: Fastest mailing address is 1543 17th Ave E, Seattle WA 98112-2808 USA. 1.206.328.1192, WilliamCalvin.org]